\documentclass{aa}
\usepackage{astron,times,graphics, amssymb}
\begin{document}
\thesaurus{06(08.01.1; 08.16.4; 08.03.1; 08.03.4; 08.12.1)}
\title{The $^{12}$C/$^{13}$C-ratio in cool carbon stars\thanks{Presented in this paper is observational data collected using the 
Swedish-ESO submillimetre telescope, La Silla, Chile,
the 20\,m telescope at Onsala Space Observatory, Chalmers Tekniska H\"ogskola, 
Sweden, and the NRAO 12\,m telescope located at Kitt Peak, USA.}}

\author{
F.~L.~Sch\"oier\inst{1} \and H.~Olofsson \inst{1}}

\institute{Stockholm Observatory, SE-133 36 Saltsj\"obaden, Sweden}

\offprints{F.~L.~Sch\"oier (fredrik@astro.su.se)}
\date{Received ; accepted}

\maketitle
\begin{abstract}
We present observations of circumstellar millimetre-wave $^{13}$CO line
emission towards a sample of 20 cool carbon stars.  Using a detailed
radiative transfer model we estimate the circumstellar
$^{12}$CO/$^{13}$CO-ratios, which we believe accurately measure the
important stellar $^{12}$C/$^{13}$C-ratios.  For those optically
bright carbon stars where it is possible, our derived
$^{12}$C/$^{13}$C-ratios are compared with the photospheric results,
obtained with different methods.  Our estimates agree well with those of
Lambert et~al. \cite*{Lambert86}.

It is shown that a straightforward determination of the
$^{12}$CO/$^{13}$CO-ratio from observed line intensity ratios is often
hampered by optical depth effects, and that a detailed radiative transfer
analysis is needed in order to determine reliable isotope ratios.
\end{abstract}


\section{Introduction}
The $^{12}$C/$^{13}$C-ratio is an important measure of stellar
evolution and nucleosynthesis.  Current theoretical models of stellar
evolution on the asymptotic giant branch (AGB) predict that stars with
``normal'' chemical composition transform into carbon stars, from
spectral type M via the intermediate states MS, S and SC, through the
dredge-up of freshly synthesized carbon in He-shell flashes (Busso et
al.\ 1999\nocite{Busso99}).  
This scenario is also confirmed by observations (Smith \&
Lambert 1985\nocite{Smith85}, 1990\nocite{Smith90}; Dominy et~al.\
1986\nocite{Dominy86}; Lambert et~al.\ 1986\nocite{Lambert86}).  Since
mainly the $^{12}$C-isotope is synthesized, there should be a coeval
evolution in the $^{12}$C/$^{13}$C-ratio from the value determined
during the first red giant evolution.  In addition, processes like hot
bottom burning will affect this ratio (Boothroyd et~al.\
1993\nocite{Boothroyd93}), as well as set an upper mass limit for
carbon stars.  Thus, an accurate estimate of the
$^{12}$C/$^{13}$C-ratio should increase our understanding of the
processes that lead to the formation of carbon stars (e.g., Forestini
\& Charbonnel 1997\nocite{Forestini97}; Wallerstein \& Knapp
1998\nocite{WK}).  The $^{12}$C/$^{13}$C-ratio is also an important
tracer of the past starformation rate and stellar mass function
(Prantzos et~al.\ 1996\nocite{Prantzos96}; Greaves \&
Holland 1997\nocite{Greaves97}).

In a classical paper, Lambert et~al.  \cite*{Lambert86} determined the
photospheric $^{12}$C/$^{13}$C-ratios of 30 optically bright carbon
stars, by fitting stellar atmosphere models to near-IR data on the
isotopomers of C$_{2}$, CO, and CN. A decade later Ohnaka \& Tsuji
(1996, 1999)\nocite{Ohnaka96}\nocite{Ohnaka99}, based on a different
method and data on the CN red system around 8000\,{\AA}, presented
$^{12}$C/$^{13}$C-ratios that, on the average, are about a factor of
two lower than those of Lambert et~al.  \cite*{Lambert86} for the same
stars.  The activity in this field was further increased with the
published results of Abia \& Isern \cite*{Abia97}.  They derived
$^{12}$C/$^{13}$C-ratios of 44 carbon stars, using the CN red system,
which fell in between the results obtained by Lambert et~al.
\cite*{Lambert86} and Ohnaka \& Tsuji \cite*{Ohnaka96}.  This is
somewhat surprising since they used the same model atmospheres as
Lambert et~al.  \cite*{Lambert86}, suggesting that the derived
$^{12}$C/$^{13}$C-ratios depend on the spectral features used.  It is
difficult to identify the main reason for the different results
obtained by Lambert et~al.  \cite*{Lambert86} and Ohnaka \& Tsuji
(1996, 1999)\nocite{Ohnaka96}\nocite{Ohnaka99}.  The former used
high-resolution near-infrared data, while the latter used data
obtained closer to optical wavelengths.  Also the atmospheric models
differ.  de~Laverny \& Gustafsson \cite*{deLaverny98} have
investigated the method used by Ohnaka \& Tsuji \cite*{Ohnaka96} and
found that it is very sensitive to model parameters and blends.  Their
conclusion is that the larger $^{12}$C/$^{13}$C-ratios determined by
Lambert et~al.  \cite*{Lambert86} are more reliable, since this
analysis is rather insensitive to the adopted model parameters, and
also the effect of blends is less severe, but the discussion continues
(Ohnaka \& Tsuji 1998\nocite{Ohnaka98}; 
de Laverny \& Gustafsson 1999\nocite{deLaverny99}).
Recently, Ohnaka et~al.  \cite*{Ohnaka00} have revised the values for
three of the stars, which were originally published in Ohnaka \& Tsuji
\cite*{Ohnaka96}.  Using new model atmospheres, they obtain
$^{12}$C/$^{13}$C-ratios that are larger by about 40\%, i.e., closer
to those estimated by Lambert et~al.  \cite*{Lambert86}.

To shed light on this disturbing controversy, we have performed independent
estimates of the $^{12}$C/$^{13}$C-ratio, using CO radio line emission from
the circumstellar envelopes (CSEs), for a sample of carbon stars
showing large discrepancies between the sets of photospheric
estimates.  Due to the weakness of the circumstellar $^{13}$CO lines,
and the difficulties in the interpretation of the circumstellar
emission, only a few attempts have been made to determine the
$^{12}$CO/$^{13}$CO ratio in the CSEs of carbon stars
(e.g., Knapp \& Chang 1985\nocite{Knapp85};
Sopka et~al.\ 1989\nocite{Sopka89};
Greaves \& Holland 1997\nocite{Greaves97}).
In this paper we present new observational
results, as well as a detailed modelling of circumstellar
$^{12}$CO and $^{13}$CO radio line emission.

\begin{table}
  \caption{Source list}
  \label{sourcelist}
  \resizebox{\hsize}{!}{
  \begin{tabular}{lrrccrc} \hline
  \noalign{\smallskip}
  \multicolumn{1}{c}{Source}    & 
  \multicolumn{1}{c}{$\alpha$}    & 
  \multicolumn{1}{c}{$\delta$}    &  
  \multicolumn{1}{c}{$^{12}{\rm C}/^{13}{\rm C}\,^1$} &  
  \multicolumn{1}{c}{$^{12}{\rm C}/^{13}{\rm C}\,^2$}\\
  &\multicolumn{1}{c}{[B1950.0]} &\multicolumn{1}{c}{[B1950.0]} & 
  & &\\
  \noalign{\smallskip}
  \hline
  \noalign{\smallskip}
   \object{R Scl}  & 01:24:40.0 & $-$32:48:07 &          19\phantom{.0} &\phantom{0}9\phantom{.0} \\
   \object{R Lep}  & 04:57:19.6 & $-$14:52:49 &          62\phantom{.0} &             \\
   \object{W Ori}  & 05:02:48.7 &    01:06:37 &          79\phantom{.0} & 26\phantom{.0}          \\
   \object{UU Aur} & 06:33:06.6 &    38:29:16 &          52\phantom{.0} & 19\phantom{.0}          \\
   \object{X Cnc}  & 08:52:34.0 &    17:25:22 &          52\phantom{.0} & 22\phantom{.0}  	     \\
   \object{U Hya}  & 10:35:05.1 & $-$13:07:27 &          32\phantom{.0} & 17\phantom{.0}          \\
   \object{V Hya}  & 10:49:11.3 & $-$20:59:05 &          69\phantom{.0} &             \\
   \object{Y CVn}  & 12:42:47.1 &    45:42:48 &\phantom{0}3.5           & \phantom{0}2.0         \\
   \object{RY Dra} & 12:54:28.2 &    66:15:53 &\phantom{0}3.6           & \phantom{0}1.9         \\
   \object{T Lyr}  & 18:30:36.1 &    36:57:38 &\phantom{0}3.2           &             \\
   \object{S Sct}  & 18:47:37.1 & $-$07:57:59 &          45\phantom{.0} & 14\phantom{.0}          \\
   \object{V Aql}  & 19:01:44.0 & $-$05:45:38 &          82\phantom{.0} & 66\phantom{.0}          \\
   \object{UX Dra} & 19:23:22.4 &    76:27:42 &          32\phantom{.0} &             \\
   \object{AQ Sgr} & 19:31:27.1 & $-$16:29:03 &          52\phantom{.0} & 35\phantom{.0}          \\
   \object{TX Psc} & 23:43:50.1 &    03:12:34 &          43\phantom{.0} & 22\phantom{.0}          \\
  \noalign{\smallskip}
  \hline
  \noalign{\smallskip}
  \noalign{$^1$ Lambert et al.\ \cite*{Lambert86}; 
           $^2$ Ohnaka \& Tsuji (1996, 1999)\nocite{Ohnaka96}\nocite{Ohnaka99}.}
  \end{tabular}}
\end{table}

\section{The observations}
We have selected a sample of 11 optically bright carbon stars for which the
photospheric $^{12}$C/$^{13}$C-ratios have been determined by Lambert et~al.
\cite*{Lambert86} and Ohnaka \& Tsuji \cite*{Ohnaka96}.  This
sample is presented in Table~\ref{sourcelist}, where we have also included
\object{S Sct}, which was observed by Bergman et~al. \cite*{Bergman93}
and Olofsson el~al. \cite*{OBEG}, as well as the three J-type stars
 (i.e., stars with $^{12}$C/$^{13}$C-ratios $\sim$3)
\object{Y~CVn},
\object{RY~Dra}, and \object{T~Lyr} (see Table~\ref{coresult} for references).
These 15 stars were also included in the major survey of
circumstellar molecular line emission by Olofsson et~al. (1993a,b).

The $^{13}$CO observations were carried out in
October 1998 using the Swedish-ESO submillimetre telescope (SEST)
located on La Silla, Chile, and in March 1999 using the Onsala 20\,m
telescope (OSO), Sweden.  At SEST, a dual channel, heterodyne SIS
receiver was used to simultaneously observe at 110\,GHz (the
$J$$=$$1$$\rightarrow$$0$ line) and 220\,GHz (the
$J$$=$$2$$\rightarrow$$1$ line).  The single sideband temperatures of
the receiver are about 110\,K and 150\,K at 110\,GHz and 220\,GHz,
respectively.  All the available acousto-optical spectrometers were
used as backends.  The two wideband (1\,GHz), low resolution
spectrometers (LRSs) were used to cover both lines.  The third, narrow
band (86\,MHz), high resolution spectrometer (HRS) was used at
220\,GHz, since the $J$$=$$2$$\rightarrow$$1$ line is usually stronger
than the lowest rotational line in CSEs. 
The LRS had 1440 channels separated by 0.7\,MHz whereas the HRS used 
2000 channels separated by 42\,kHz.

At OSO, an SIS
receiver, with a single sideband temperature of about 100\,K at
110\,GHz, was used for the observations.  As backends, two filterbanks
with bandwidths of 512\,MHz (MUL~A) and 64\,MHz (MUL~B) were used.
The MUL~A used 512 channels separated by 1\,MHz
and the MUL~B filterbank used 256 channels with a separation of 250\,kHz.

\begin{table}
  \caption{Observational $^{13}$CO results. A colon (:) marks an uncertain
  value due to a low signal to noise ratio (\object{UU~Aur}) or contamination
  by interstellar lines (\object{UX~Dra}).}
  \label{obs}
  \resizebox{\hsize}{!}{
  \begin{tabular}{llccccc} \hline
  \noalign{\smallskip}
  \multicolumn{1}{c}{Source}    & 
  \multicolumn{1}{c}{Tel.}    & 
  \multicolumn{1}{c}{Trans.}    & 
  \multicolumn{1}{c}{$T_{\mathrm{mb}}$}    & 
  \multicolumn{1}{c}{$I_{\mathrm{mb}}$}    &  
  \multicolumn{1}{c}{$v_*$}    & 
  \multicolumn{1}{c}{$v_\mathrm{e}$} \\
  & 
  &
  &
  \multicolumn{1}{c}{[K]} &
  \multicolumn{1}{c}{[K\,km\,s$^{-1}$]} &
  \multicolumn{1}{c}{[km\,s$^{-1}$]} &
  \multicolumn{1}{c}{[km\,s$^{-1}$]} \\
  \noalign{\smallskip}
  \hline
  \noalign{\smallskip}
   \object{R Scl}  & SEST & 1$-$0 & 0.055\phantom{:}  &\phantom{$<$}1.9\phantom{0:} & $-$19.0\phantom{:}          & 15.1\phantom{:} \\
                   & SEST & 2$-$1 & 0.11\phantom{0:}  &\phantom{$<$}4.1\phantom{0:} & $-$19.0\phantom{:}          & 15.2\phantom{:} \\
   \object{R Lep}  & SEST & 1$-$0 &		      & $<$0.20\phantom{:}	    &		                  &         \\
                   & SEST & 2$-$1 & 0.015\phantom{:}  &\phantom{$<$}0.47\phantom{:} &\phantom{$-$}10.1\phantom{:} & 16.7\phantom{:} \\
   \object{W Ori}  & SEST & 1$-$0 &		      & $<$0.18\phantom{:}	    &			          &      \\
                   & SEST & 2$-$1 &		      & $<$0.25\phantom{:}	    &			          &      \\
   \object{UU Aur} & OSO  & 1$-$0 & 0.016:	      &\phantom{$<$}0.17:	    &\phantom{$-$0}8.4:           & 10.2: \\
   \object{X Cnc}  & SEST & 1$-$0 &		      & $<$0.28\phantom{:}	    &			          &      \\
                   & SEST & 2$-$1 &		      & $<$0.27\phantom{:}	    &			          &       \\
   \object{U Hya}  & SEST & 2$-$1 & 0.066\phantom{:}  &\phantom{$<$}0.77\phantom{:} & $-$31.1\phantom{:} 	  &\phantom{0}5.8\phantom{:} \\
   \object{V Hya}  & SEST & 1$-$0 & 0.031\phantom{:}  &\phantom{$<$}0.67\phantom{:} & $-$14.3\phantom{:} 	  & 10.7\phantom{:} \\
                   & SEST & 2$-$1 & 0.032\phantom{:}  &\phantom{$<$}1.1\phantom{0:} & $-$17.0\phantom{:}          & 13.0\phantom{:} \\
   \object{Y CVn}  & OSO  & 1$-$0 & 0.24\phantom{0:}  &\phantom{$<$}3.4\phantom{0:} &\phantom{$-$}21.8\phantom{:} &\phantom{0}8.8\phantom{:} \\
   \object{V Aql}  & SEST & 1$-$0 &		      & $<$0.23\phantom{:}  	    &	       	                  &      \\
                   & SEST & 2$-$1 &		      & $<$0.29\phantom{:}          &		                  &      \\
   \object{UX Dra} & OSO  & 1$-$0 & 0.019:            &\phantom{$<$}0.14:	    &\phantom{$-$}15.5:           &\phantom{0}4.5: \\
   \object{AQ Sgr} & SEST & 1$-$0 &		      & $<$0.12\phantom{:}          &		      &      \\
                   & SEST & 2$-$1 &		      & $<$0.27\phantom{:}  	    &		      &      \\
   \object{TX Psc} & SEST & 1$-$0 & 0.009\phantom{:}  &\phantom{$<$}0.12\phantom{:} &\phantom{$-$}13.5\phantom{:} &\phantom{0}8.1\phantom{:}  \\
                   & SEST & 2$-$1 & 0.023\phantom{:}  &\phantom{$<$}0.22\phantom{:} &\phantom{$-$}11.8\phantom{:} &\phantom{0}8.3\phantom{:}  \\
  \noalign{\smallskip}
  \hline
  \noalign{\smallskip}
  \end{tabular}}
\end{table}

\begin{figure*}
  \resizebox{\hsize}{!}{\includegraphics{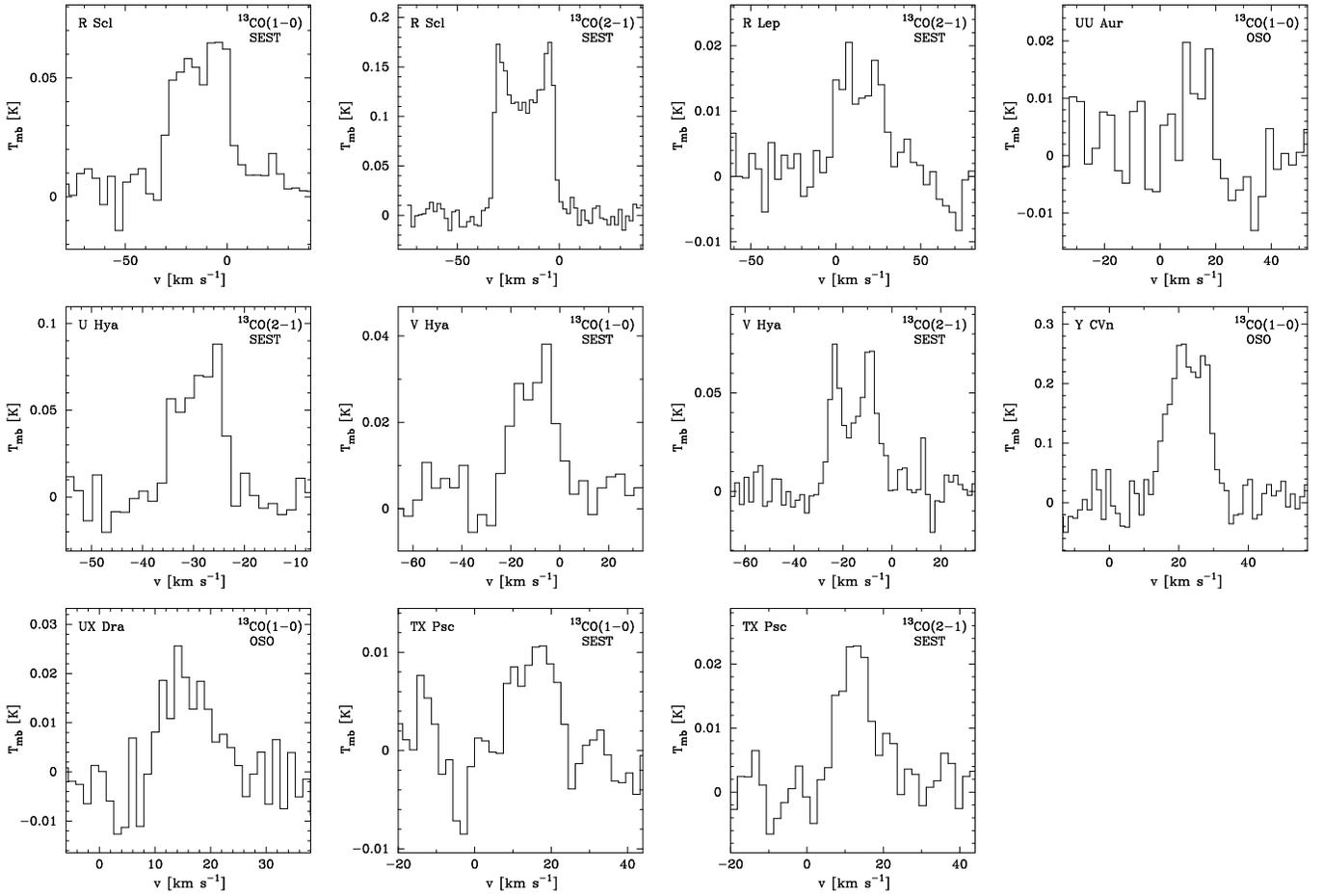}}
  \caption{New observations of circumstellar $^{13}$CO line emission
  including first detections (see text for details).  The
  line observed and the telescope used are shown in the upper right
  corner of each panel}
  \label{spectra}
\end{figure*}

The observations were made in a dual beamswitch mode, where the source
is alternately placed in the signal and the reference beam, using a
beam throw of about 12$\arcmin$ at SEST and about 11$\arcmin$ at
OSO. This method produces two spectra that are subtracted from each
other, which results in very flat baselines, i.e., most of the frequency 
dependant response of the system is removed.
The intensity scales are
given in main beam brightness temperature, $T_{\mathrm{mb}}$$=$$T_{\mathrm
A}^{\star}/\eta_{\mathrm{mb}}$, where $T_{\mathrm A}^{\star}$ is the
antenna temperature corrected for atmospheric attenuation using the
chopper wheel method, and $\eta_{\mathrm{mb}}$ is the main beam
efficiency.  For the SEST $\eta_{\mathrm{mb}}$ is 0.7 at 110\,GHz and
0.5 at 220\,GHz.  At OSO $\eta_{\mathrm{mb}}$$=$0.5 at 110\,GHz.  The
beamsizes of the SEST are 45$\arcsec$ and 24$\arcsec$ at 110\,GHz and
220\,GHz, respectively.  At OSO the beam is 34$\arcsec$ at 110\,GHz.
Regular pointing checks were made on strong SiO masers and the
pointing was usually better than $\pm 3\arcsec$ for both telescopes.
The uncertainty in the intensity scale is estimated to be about
$\pm$20\%.  The observational results are summarized in Table~\ref{obs}
and the detections of $^{13}$CO line emission are shown in
Fig.~\ref{spectra}. These observations include first detections of
circumstellar
$^{13}$CO emission towards \object{R~Lep}, \object{UU~Aur}, \object{UX~Dra},
and \object{TX~Psc}. The line parameters, i.e., the main beam
brightness temperature at the line centre ($T_{\mathrm{mb}}$), the line centre
velocity ($v_{*}$), and half the full line width ($v_{\mathrm e}$),
are obtained by fitting the data with an artificial line profile (Olofsson et
al.\ 1993a)
\begin{equation}
T(v) = T_{\mathrm{mb}} \left[ 1 - \left( \frac{v-v_*}{v_{\mathrm e}} \right) ^2
\right] ^{\gamma/2},
\end{equation}
where $\gamma$ is a parameter describing the shape of the line.
The integrated intensity ($I_{\mathrm{mb}}$) is obtained by integrating
the emission between $v_*$$\pm$$v_{\mathrm e}$.  The upper limits are
obtained as $T_{\rm pp}2v_{\rm e}$, where $T_{\rm pp}$ is the
peak-to-peak noise obtained from spectra where the velocity resolution
has been degraded to twice the expansion velocity.  The corresponding
$^{12}$CO lines were also observed, simultaneously with the $^{13}$CO
lines, for all the sample stars.  Spectra of these lines have been
published in Olofsson et~al.\ (1993a).

To complement these new observations we have collected additional
$^{13}$CO data and, in particular, $^{12}$CO data on the sample stars,
as well as for an additional seven stars for comparison, from various
sources and included them in the modelling (see Table~\ref{coresult}).

\begin{table*}
  \caption{Input parameters and derived $^{12}$CO/$^{13}$CO-ratios from the radiative transfer
  calculations. A colon (:) marks an uncertain $^{12}$CO/$^{13}$CO-ratio 
  estimate that is based on observations of only one $^{13}$CO transition.}
  \label{input}
  \begin{tabular}{lcccccccccc} \hline
  \noalign{\smallskip}
  \multicolumn{1}{c}{Source}    & 
  \multicolumn{1}{c}{$D$}    & 
  \multicolumn{1}{c}{$L$}    & 
  \multicolumn{1}{c}{$T_*$}    & 
  \multicolumn{1}{c}{$T_{\mathrm d}$}    & 
   \multicolumn{1}{c}{$L_{\mathrm d}/L_*$}    & 
  \multicolumn{1}{c}{$\dot{M}$}    & 
  \multicolumn{1}{c}{$v_{\mathrm{e}}$}    &
  \multicolumn{1}{c}{$r_{\mathrm p}$}    & 
  \multicolumn{1}{c}{$^{12}{\rm CO}/^{13}{\rm CO}$} &
 
\multicolumn{1}{c}{$\tau$[$^{12}$CO($2$$\rightarrow$$1$)]$^{\mathrm b}$} 
  \\
  &
   \multicolumn{1}{c}{[pc]} & 
   \multicolumn{1}{c}{[L$_{\sun}$]} & 
   \multicolumn{1}{c}{[K]} & 
   \multicolumn{1}{c}{[K]} & 
   &
   \multicolumn{1}{c}{[M$_{\sun}$/yr]} & 
   \multicolumn{1}{c}{[km\,s$^{-1}$]} &
   \multicolumn{1}{c}{[cm]} & 
  \\
  \noalign{\smallskip}
  \hline
  \noalign{\smallskip}
\object{R Scl}$^{\mathrm a}$ & 360$^2$ &\phantom{0}5500 & 2000 &     &      &\phantom{0.}4$\times$10$^{-6}$ & 16.5	   &7.0$\times$10$^{16}$&\phantom{$>$}20\phantom{:.0} & 1.5\phantom{0}\\
   \object{V384 Per}         & 560$^2$ &\phantom{0}8100 & 1900 & 890 &10.0\phantom{0}     &3.5$\times$10$^{-6}$ & 15.0	   & 1.4$\times$10$^{17}$ &\phantom{$>$}35\phantom{:.0} & 2.4\phantom{0}  \\
   \object{R Lep}            & 250$^1$ &\phantom{0}4000 & 1900 & 820 & \phantom{0}0.14 &\phantom{0.}7$\times$10$^{-7}$ &18.0
      & 5.3$\times$10$^{16}$&\phantom{$>$}90:\phantom{.0} &0.90 \\
   \object{W Ori}            & 220$^1$ &\phantom{0}2600 & 2200 &     &      &\phantom{0.}7$\times$10$^{-8}$ & 11.0	  
   &1.9$\times$10$^{16}$ & $>$30\phantom{:.0} & 0.28 \\
   \object{UU Aur}           & 260$^2$ &\phantom{0}6900 & 2500 &     &      &3.5$\times$10$^{-7}$ & 11.0 &
   4.2$\times$10$^{16}$ &\phantom{$>$}45:\phantom{.0}  &1.1\phantom{0}\\
   \object{X Cnc}            & 280$^2$ &\phantom{0}2800 & 2200 &     &      &1.0$\times$10$^{-7}$
   &\phantom{0}7.0&2.4$\times$10$^{16}$ & $>$30\phantom{:.0}  & 1.3\phantom{0}\\
   \object{CW Leo}           & 120$^2$ &\phantom{0}9600 &      & 510 &      &1.5$\times$10$^{-5}$           & 14.5	  
   &3.7$\times$10$^{17}$ &\phantom{$>$}50\phantom{:.0} &  5.4\phantom{0}\\
   \object{RW LMi}           & 440$^2$ &\phantom{0}9700 & 1300 & 510 & \phantom{0}6.7\phantom{0}
   &\phantom{0.}6$\times$10$^{-6}$ & 17.0	   & 1.9$\times$10$^{17}$ &\phantom{$>$}35\phantom{:.0}  & 1.7\phantom{0}\\
   \object{U Hya}            & 160$^1$ &\phantom{0}2500 & 2400 &     &      &1.4$\times$10$^{-7}$ &\phantom{0}7.0&
   2.9$\times$10$^{16}$ &\phantom{$>$}40\phantom{:.0} &1.4\phantom{0} \\
   \object{Y CVn}            & 220$^1$ &\phantom{0}4400 & 2200 &     &      &1.5$\times$10$^{-7}$           &\phantom{0}8.5&
   3.1$\times$10$^{16}$ &\phantom{$>$0}2.5\phantom{:} & 0.87\\
   \object{RY Dra}           & 490$^1$ &          10000 & 2500 &     &      &\phantom{0.}3$\times$10$^{-7}$ & 10.0	  &4.0$\times$10$^{16}$ &\phantom{$>$0}3.0: & 0.40 \\
   \object{IRAS 15194-5115}  & 600$^2$ &\phantom{0}8800 &\phantom{0}930 & 480 & \phantom{0}2.2\phantom{0}
   &1.0$\times$10$^{-5}$ & 21.5	   & 3.2$\times$10$^{17}$ &\phantom{$>$0}5.5\phantom{:} & 1.4\phantom{0} \\
   \object{T Lyr}            & 340$^3$ &\phantom{0}4000 & 2000 &     &      &\phantom{0.}7$\times$10$^{-8}$ & 11.5	  
   &1.8$\times$10$^{16}$ &\phantom{$>$0}3.0: & 0.14 \\
\object{S Sct}$^{\mathrm a}$ & 400$^1$ &\phantom{0}4900 & 2400 &     &      &\phantom{0.}4$\times$10$^{-5}$ & 16.5	   &
4.0$\times$10$^{17}$ &\phantom{$>$}35:\phantom{.0} &0.90 \\
   \object{V Aql}            & 370$^1$ &\phantom{0}6500 & 2100 &     &      &\phantom{0.}3$\times$10$^{-7}$ &\phantom{0}8.5&
   4.2$\times$10$^{16}$ &	     $>$70\phantom{:.0} & 1.4\phantom{0}  \\
   \object{UX Dra}           & 310$^3$ &\phantom{0}4000 & 2600 &     &      &1.6$\times$10$^{-7}$ &\phantom{0}4.0&
   4.0$\times$10$^{16}$&\phantom{$>$}25:\phantom{.0} & 2.9\phantom{0} \\
   \object{AQ Sgr}           & 420$^3$ &\phantom{0}4000 & 2700 &     &      &2.5$\times$10$^{-7}$                  & 10.0
      & 3.6$\times$10$^{16}$ &	     $>$40\phantom{:.0}  & 0.93\\
   \object{V Cyg}            & 370$^2$ &\phantom{0}6200 & 1500 & 860 & \phantom{0}0.84 & 1.2$\times$10$^{-6}$              
   & 11.5	   & 8.5$\times$10$^{16}$ &\phantom{$>$}20\phantom{:.0} &1.1\phantom{0} \\
   \object{RV Aqr}           & 670$^2$ &\phantom{0}6800 & 1300 & 620 & \phantom{0}0.46 & 2.5$\times$10$^{-6}$               
     & 16.0	   & 1.1$\times$10$^{17}$ &\phantom{$>$}35:\phantom{.0}  &1.5\phantom{0}\\
   \object{LP And}           & 630$^2$ &\phantom{0}9400 & 1100 & 610 & \phantom{0}6.6\phantom{0} &1.5$\times$10$^{-5}$ &
   14.0	   & 4.0$\times$10$^{17}$ &\phantom{$>$}55\phantom{:.0} & 7.3\phantom{0} \\
  \noalign{\smallskip}
  \hline
  \noalign{\smallskip}
  \noalign{$^{\mathrm a}$ Detached CSE; $^{\mathrm b}$ The maximum tangential optical depth 
  in the $^{12}$CO($J$$=$$2$$\rightarrow$$1$) derived from the model.} 
  \noalign{\smallskip}
  \noalign{
  $^1$ Hipparcos; $^2$ Period-luminosity relation 
  (Groenewegen \& Whitelock 1996\nocite{Groenewegen96}); 
  $^3$ $L$$=$4000\,L$_{\sun}$ assumed.}
  \end{tabular}
\end{table*}

\begin{table*}
\caption[ ]{CO modelling results compared with observations.}
\begin{flushleft}
\begin{tabular}{llcccccccccc}
\hline
\noalign{\smallskip}

& & &
 \multicolumn{3}{c}{$^{12}$CO} & &
 \multicolumn{3}{c}{$^{13}$CO} 
\\
\cline{4-6}
\cline{8-10}
\\
 \multicolumn{1}{c}{Source} & 
 \multicolumn{1}{c}{Tel.} & 
 \multicolumn{1}{c}{Trans.}& 
 \multicolumn{1}{c}{$I_{\mathrm{obs}}$} &
 \multicolumn{1}{c}{$I_{\mathrm{mod}}$} & 
 \multicolumn{1}{c}{Ref.} & &
 \multicolumn{1}{c}{$I_{\mathrm{obs}}$} &
 \multicolumn{1}{c}{$I_{\mathrm{mod}}$} &
 \multicolumn{1}{c}{Ref.} & &
 \multicolumn{1}{c}{$I$($^{12}$CO)/$I$($^{13}$CO)$^{\mathrm a}$}
 \\
  &
  &
  &
 \multicolumn{1}{c}{[K\,km\,s$^{-1}$]} &
 \multicolumn{1}{c}{[K\,km\,s$^{-1}$]} & & &
 \multicolumn{1}{c}{[K\,km\,s$^{-1}$]} &
 \multicolumn{1}{c}{[K\,km\,s$^{-1}$]} & & &
 \\
\noalign{\smallskip}
\hline
\noalign{\smallskip}
\object{R Scl}   & SEST  & 1$-$0 & \phantom{00}26.1 & \phantom{00}29.6 & 4 & & \phantom{$<$00}1.9\phantom{0} & \phantom{00}2.3\phantom{0}& 1 & &\phantom{$<$}12   \\
                 & IRAM  & 1$-$0 & \phantom{00}76.2 & \phantom{00}79.2 & 4 & & &  &    \\
		 & SEST  & 2$-$1 & \phantom{00}47.8 & \phantom{00}49.8 & 4 & & \phantom{$<$00}4.2\phantom{0} & \phantom{00}4.0\phantom{0}& 1 & &\phantom{$<$}10   \\
                 & JCMT  & 2$-$1 & \phantom{00}50.6 & \phantom{00}54.8 & 7 & &   \\
		 & IRAM  & 2$-$1 & \phantom{00}67.7 & \phantom{00}72.0 & 4 & & &  &    \\
		 & SEST  & 3$-$2 & \phantom{00}63.7 & \phantom{00}28.0 & 6 & & &  &    \\
		 & JCMT  & 3$-$2 & \phantom{00}62.9 & \phantom{00}30.5 & 7 & & &  &    \\
\object{V384 Per}& NRAO  & 1$-$0 & \phantom{000}7.8 & \phantom{00}10.1 & 2 & & &  &    \\
                 & OSO   & 1$-$0 & \phantom{00}25.5 & \phantom{00}25.1 & 4 & & \phantom{$<$00}2.1\phantom{0} & \phantom{00}2.1\phantom{0}& 5 & &\phantom{$<$}11 \\
                 & IRAM  & 1$-$0 & \phantom{00}52.7 & \phantom{00}48.4 & 4 & & &  &    \\
		 & JCMT  & 2$-$1 & \phantom{00}37.3 & \phantom{00}35.2 & 7 & & &  &    \\
		 & IRAM  & 2$-$1 & \phantom{00}83.6 & \phantom{0}102.0 & 4 & & &  &    \\
		 & JCMT  & 3$-$2 & \phantom{00}44.5 & \phantom{00}42.5 & 7 & & \phantom{$<$00}4.4\phantom{0} & \phantom{00}4.3\phantom{0}& 7 & &\phantom{$<$0}9  \\
\object{R Lep}   & SEST  & 1$-$0 & \phantom{000}6.2 & \phantom{000}6.7 & 4 & & \phantom{00}$<$0.20  &\phantom{00}0.07  & 1 &
&$>$33 \\
                 & IRAM  & 1$-$0 & \phantom{00}31.7 & \phantom{00}24.3 & 4 & & &  &    \\
		 & SEST  & 2$-$1 & \phantom{00}18.1 & \phantom{00}17.4 & 4 & & \phantom{$<$00}0.5\phantom{0}  &
		 \phantom{00}0.5\phantom{0} & 1 & &\phantom{$<$}31 \\
                 & IRAM  & 2$-$1 & \phantom{00}56.4 & \phantom{00}64.2 & 4 & & &  &    \\
		 & SEST  & 3$-$2 & \phantom{00}12.5 & \phantom{00}21.6 & 2 & & &  &    \\
		 & JCMT  & 3$-$2 & \phantom{00}28.1 & \phantom{00}27.9 & 7 & & &  &    \\
\object{W Ori}   & SEST  & 1$-$0 & \phantom{000}1.2 & \phantom{000}1.3 & 2 & & \phantom{00}$<$0.18  & \phantom{00}0.02 & 1 & &\phantom{0}$>$6 \\
                 & OSO   & 1$-$0 & \phantom{000}2.7 & \phantom{000}2.4 & 4 & & &  &    \\
		 & IRAM  & 1$-$0 & \phantom{000}5.0 & \phantom{000}5.1 & 4 & & &  &    \\
		 & SEST  & 2$-$1 & \phantom{000}4.9 & \phantom{000}4.3 & 2 & & \phantom{00}$<$0.25  & \phantom{00}0.25  & 1
		 & &$>$17 \\
                 & IRAM  & 2$-$1 & \phantom{00}19.0 & \phantom{00}16.9 & 4 & & &  &    \\
		 & SEST  & 3$-$2 & \phantom{000}4.8 & \phantom{000}5.4 & 2 & & &  &    \\
\object{UU Aur}  & OSO   & 1$-$0 & \phantom{000}7.9 & \phantom{000}8.2 & 4 & & \phantom{$<$00}0.17 & \phantom{00}0.17 & 1 & &\phantom{$<$}40 \\
                 & IRAM  & 1$-$0 & \phantom{00}18.8 & \phantom{00}16.9 & 4 & & &  &    \\
		 & JCMT  & 2$-$1 & \phantom{00}16.8 & \phantom{00}14.8 & 7 & & &  &    \\
		 & IRAM  & 2$-$1 & \phantom{00}39.0 & \phantom{00}46.2 & 4 & & &  &    \\
\object{X Cnc}   & NRAO  & 1$-$0 & \phantom{000}0.7 & \phantom{000}0.7 & 2 & &  	&	   &	     \\
                 & SEST  & 1$-$0 & \phantom{000}1.5 & \phantom{000}1.1 & 2 & & \phantom{00}$<$0.28  & \phantom{00}0.03 & 1& &\phantom{0}$>$5\\
		 & OSO   & 1$-$0 & \phantom{000}1.7 & \phantom{000}2.0 & 4 & &     &	      &    \\
                 & SEST  & 2$-$1 & \phantom{000}3.2 & \phantom{000}2.9 & 2 & & \phantom{00}$<$0.27  & \phantom{00}0.27 & 1& &$>$10\\
		 & IRAM  & 2$-$1 & \phantom{00}11.1 & \phantom{00}11.8 & 4 & &         &	  &    \\
\object{CW Leo}  & NRAO  & 1$-$0 & \phantom{0}170.8 & \phantom{0}257.1 & 2 & & \phantom{$<$0}16.9\phantom{0} &\phantom{0}22.1\phantom{0} & 3 && \phantom{00}9 \\
                 & SEST  & 1$-$0 & \phantom{0}288.1 & \phantom{0}305.3 & 4 & & \phantom{$<$0}24.3\phantom{0}
		 &\phantom{0}26.8\phantom{0} & 5 && \phantom{0}11 \\
		 & OSO   & 1$-$0 & \phantom{0}422.0 & \phantom{0}391.7 & 4 & & \phantom{$<$0}31.1\phantom{0}
		 &\phantom{0}32.7\phantom{0} & 3 && \phantom{0}12 \\
                 & SEST  & 2$-$1 & \phantom{0}487.3 & \phantom{0}580.5 & 4 & & \phantom{$<$0}64.9\phantom{0}
		 &\phantom{0}73.9\phantom{0} & 3 && \phantom{00}7 \\
                 & JCMT  & 2$-$1 & \phantom{0}731.8 & \phantom{0}632.2 & 7 & & \phantom{$<$0}76.9\phantom{0} &\phantom{0}79.9\phantom{0} & 7 && \phantom{00}8 \\
		 & IRAM  & 2$-$1 & 	     1057.7 &   	1116.1 & 4 & &  	& \\
                 & SEST  & 3$-$2 & \phantom{0}854.8 & \phantom{0}774.0 & 2 & &  	& \\
		 & JCMT  & 3$-$2 & 	     1066.3 & \phantom{0}883.7 & 7 & & \phantom{$<$}157.3\phantom{0} &130.4\phantom{0} & 7 &&  \phantom{00}6\\
		 & JCMT  & 4$-$3 & 	     1227.8 &   	1052.0 & 7 & &  	& \\
\object{RW LMi}  & NRAO  & 1$-$0 & \phantom{00}36.4 & \phantom{00}37.0 & 2 & & \phantom{$<$00}2.4\phantom{0} &\phantom{00}2.4\phantom{0} & 3 && \phantom{0}13 \\
		 & SEST  & 1$-$0 & \phantom{00}54.6 & \phantom{00}51.9 & 4 & &  	& \\
		 & OSO   & 1$-$0 & \phantom{00}87.0 & \phantom{00}83.8 & 4 & & \phantom{$<$00}5.6\phantom{0} &\phantom{00}5.3\phantom{0} & 5 && \phantom{0}14 \\
		 & SEST  & 2$-$1 & \phantom{0}105.4 & \phantom{0}128.4 & 4 & &  	& \\
		 & JCMT  & 2$-$1 & \phantom{0}163.2 & \phantom{0}147.4 & 7 & & \phantom{$<$0}10.3\phantom{0} &\phantom{0}11.9\phantom{0} & 7 && \phantom{0}14 \\
		 & JCMT  & 3$-$2 & \phantom{0}245.3 & \phantom{0}208.6 & 7 & & \phantom{$<$0}13.8\phantom{0} &\phantom{0}16.4\phantom{0} & 7 && \phantom{0}15 \\
                 & JCMT  & 4$-$3 & \phantom{0}243.1 & \phantom{0}224.3 & 7 & &  	 & \\
\object{U Hya}   & SEST  & 1$-$0 & \phantom{00}5.4  & \phantom{00}5.4  & 4 & & \phantom{$<$0}0.20 & 0.13 & 5&& \phantom{$>$}23\phantom{.0}\\
                 & SEST  & 2$-$1 & \phantom{0}13.8  & \phantom{0}14.1  & 4 & & \phantom{$<$0}0.8\phantom{0} &  1.0\phantom{0} & 1 && \phantom{$>$}15\phantom{.0}\\
		 & JCMT  & 2$-$1 & \phantom{0}20.2  & \phantom{0}16.6  & 7 & &      &	    &	 \\
		 & IRAM  & 2$-$1 & \phantom{0}48.8  & \phantom{0}48.7  & 4 & &      &	    &	 \\
\noalign{\smallskip}
\hline
\noalign{\smallskip}
\end{tabular}
\end{flushleft}
\label{coresult}
\end{table*}

\addtocounter{table}{-1}
\begin{table*}[!]
\caption[ ]{continued}
\begin{flushleft}
\begin{tabular}{llcccccccccc}
\hline
\noalign{\smallskip}

& & &
 \multicolumn{3}{c}{$^{12}$CO} & &
 \multicolumn{3}{c}{$^{13}$CO} 
\\
\cline{4-6}
\cline{8-10}
\\
 \multicolumn{1}{c}{Source} & 
 \multicolumn{1}{c}{Tel.} & 
 \multicolumn{1}{c}{Trans.}& 
 \multicolumn{1}{c}{$I_{\mathrm{obs}}$} &
 \multicolumn{1}{c}{$I_{\mathrm{mod}}$} & 
 \multicolumn{1}{c}{Ref.} & &
 \multicolumn{1}{c}{$I_{\mathrm{obs}}$} &
 \multicolumn{1}{c}{$I_{\mathrm{mod}}$} &
 \multicolumn{1}{c}{Ref.} & &
 \multicolumn{1}{c}{$I$($^{12}$CO)/$I$($^{13}$CO)$^{\mathrm a}$}
 \\
  &
  &
  &
 \multicolumn{1}{c}{[K\,km\,s$^{-1}$]} &
 \multicolumn{1}{c}{[K\,km\,s$^{-1}$]} & & &
 \multicolumn{1}{c}{[K\,km\,s$^{-1}$]} &
 \multicolumn{1}{c}{[K\,km\,s$^{-1}$]} & & &
 \\
\noalign{\smallskip}
\hline
\noalign{\smallskip}
\object{Y CVn}   & NRAO  & 1$-$0 & 	            &                  &   & & \phantom{$<$0}1.0\phantom{0} &  1.0\phantom{0} & 1  \\
                 & OSO   & 1$-$0 & \phantom{00}4.5  & \phantom{00}5.4  & \phantom{0}4 & & \phantom{$<$0}3.4\phantom{0} & 2.4\phantom{0} & 1&& \phantom{$>$0}1.2  \\
                 & IRAM  & 1$-$0 & \phantom{0}10.3  & \phantom{0}11.2  & \phantom{0}4 & &         & &     \\
		 & JCMT  & 2$-$1 & \phantom{0}11.9  & \phantom{0}10.9  & \phantom{0}7 & &      &	    &	 \\
		 & IRAM  & 2$-$1 & \phantom{0}22.7  & \phantom{0}34.2  & \phantom{0}4 & &      &	    &	 \\
                 & JCMT  & 3$-$2 & \phantom{0}20.9  & \phantom{0}15.6  & \phantom{0}7 & & \phantom{$<$0}7.9\phantom{0} & 7.7\phantom{0} & 7 && \phantom{$>$0}2.3\\
\object{RY Dra}  & OSO   & 1$-$0 & \phantom{00}2.4  & \phantom{00}3.0  & \phantom{0}4 & &  &  &       \\
                 & IRAM  & 2$-$1 & \phantom{0}21.6  & \phantom{0}28.3  & \phantom{0}4 & &      & 	 &     \\
                 & JCMT  & 3$-$2 & \phantom{0}18.9  & \phantom{0}15.8  & \phantom{0}7 & & \phantom{$<$0}5.8\phantom{0} & 5.4\phantom{0} & 7 && \phantom{$>$0}2.8\\
\object{IRAS 15194-5115}& SEST&1$-$0&\phantom{0}53.5& \phantom{0}49.1  & \phantom{0}9 & & \phantom{$<$}14.1\phantom{0}  & \phantom{0}13.3  & 9 && \phantom{$>$0}3.3\\
                        & SEST&2$-$1&          114.0&           111.2  & \phantom{0}9 & & \phantom{$<$}30.5\phantom{0}  &\phantom{0}32.8  & 9 && \phantom{$>$0}3.3\\
                        & SEST&3$-$2&          131.5&		138.7  &           10 & &  \\
\object{T Lyr}   & OSO   & 1$-$0 & \phantom{00}0.7  & \phantom{00}1.0  & \phantom{0}4 & &      &	     &     \\
                 & IRAM  & 2$-$1 & \phantom{00}5.6  & \phantom{00}9.6  & \phantom{0}4 & &      &	     &     \\
                 & JCMT  & 3$-$2 & \phantom{00}6.4  & \phantom{00}4.3  & \phantom{0}7 & & \phantom{$<$0}1.8\phantom{0} & 2.0\phantom{0} & 7&& \phantom{$>$0}3.1\\
\object{S Sct}   & SEST  & 1$-$0 & \phantom{00}5.3  & \phantom{00}4.9  & \phantom{0}6 & & \phantom{$<$0}0.21 & 0.19 & 6  && \phantom{$>$}22\phantom{.0}\\
                 & IRAM  & 1$-$0 & \phantom{00}5.3  & \phantom{00}4.7  & \phantom{0}4 & &         & &     \\
		 & SEST  & 2$-$1 & \phantom{00}2.7  & \phantom{00}2.7  & \phantom{0}6 & &  &  &   \\
		 & JCMT  & 2$-$1 & \phantom{00}2.4  & \phantom{00}2.8  & \phantom{0}7 & &  &  &   \\
		 & IRAM  & 2$-$1 & \phantom{00}2.8  & \phantom{00}2.4  & \phantom{0}4 & &         & &     \\
		 & SEST  & 3$-$2 & \phantom{00}1.0  & \phantom{00}0.9  & \phantom{0}6 & &         & &     \\
                 & JCMT  & 3$-$2 & \phantom{00}1.0  & \phantom{00}0.9  & \phantom{0}7 & &         & &     \\
\object{V Aql}   & SEST  & 1$-$0 & \phantom{00}2.8  & \phantom{00}2.6  & \phantom{0}4 & & \phantom{0}$<$0.23 & 0.03 & 1&&$>$11\phantom{.0} \\
                 & OSO   & 1$-$0 & \phantom{00}3.2  & \phantom{00}4.8  & \phantom{0}4 & &         &	 &	\\
		 & SEST  & 2$-$1 & \phantom{00}8.1  & \phantom{00}7.0  & \phantom{0}4 & & \phantom{0}$<$0.29 & 0.28 & $1$&& $>$24\phantom{.0} \\
		 & JCMT  & 2$-$1 & \phantom{00}9.0  & \phantom{00}8.3  & \phantom{0}7 & &         &	 &	\\
		 & JCMT  & 3$-$2 & \phantom{0}11.2  & \phantom{0}10.6  & \phantom{0}7 & &         &	 &	\\

\object{UX Dra}  & OSO   & 1$-$0 & \phantom{00}2.1  & \phantom{00}1.7  & \phantom{0}4 & & \phantom{$<$0}0.14 & 0.14 & 1 && \phantom{$>$0}9\phantom{.0} \\
                 & IRAM  & 2$-$1 & \phantom{00}6.8  & \phantom{00}9.0  & \phantom{0}4 & &         & &     \\
\object{AQ Sgr}  & SEST  & 1$-$0 & \phantom{00}1.4  & \phantom{00}1.3  & \phantom{0}4 & & \phantom{0}$<$0.12 & 0.04 & 1 && $>$10\phantom{.0} \\
                 & SEST  & 2$-$1 & \phantom{00}4.1  & \phantom{00}4.0  & \phantom{0}4 & & \phantom{0}$<$0.27 & 0.25 & 1 && $>$13\phantom{.0} \\
\object{V Cyg }  & OSO   & 1$-$0 & \phantom{0}27.5  & \phantom{0}27.2  & \phantom{0}4 & & \phantom{$<$0}2.3\phantom{0} & 1.9\phantom{0} & 5 && \phantom{$>$}10\phantom{.0} \\
                 & NRAO  & 2$-$1 & \phantom{0}50.1  & \phantom{0}36.5  & \phantom{0}2 & &         & &     \\
		 & JCMT  & 2$-$1 & \phantom{0}35.9  & \phantom{0}56.2  & \phantom{0}7 & &         & &     \\
		 & JCMT  & 3$-$2 & \phantom{0}88.9  & \phantom{0}86.1  & \phantom{0}7 & & \phantom{$<$0}6.7\phantom{0} & 8.5\phantom{0} & 7 && \phantom{$>$}12\phantom{.0} \\
                 & JCMT  & 4$-$3 & 	     123.4  & \phantom{0}95.2  & \phantom{0}7 & &         & &     \\
\object{RV Aqr}  & SEST  & 1$-$0 & \phantom{00}7.5  & \phantom{00}6.9  & \phantom{0}4 & & \phantom{$<$0}0.6\phantom{0} & 0.6\phantom{0} & 5 && \phantom{$>$}11\phantom{.0} \\
                 & SEST  & 2$-$1 & \phantom{0}18.1  & \phantom{0}17.2  & \phantom{0}4 & &         & &     \\
		 & SEST  & 3$-$2 & \phantom{0}18.6  & \phantom{0}20.5  & \phantom{0}2 & &         & &     \\
\object{LP And}  & OSO   & 1$-$0 & \phantom{0}57.0  & \phantom{0}58.8  & \phantom{0}2 & & \phantom{$<$0}5.0\phantom{0} & 5.2\phantom{0} & 3 && \phantom{$>$}10\phantom{.0} \\
                 & IRAM  & 1$-$0 & \phantom{0}92.7  & \phantom{0}97.4  & \phantom{0}8 & &         & &     \\
		 & NRAO  & 2$-$1 & \phantom{0}60.7  & \phantom{0}46.5  & \phantom{0}2 & &         & &     \\
		 & JCMT  & 2$-$1 & \phantom{0}90.6  & \phantom{0}68.5  & \phantom{0}7 & & \phantom{$<$0}8.7\phantom{0} & 8.3\phantom{0} & 7 && \phantom{$>$0}9\phantom{.0} \\
                 & IRAM  & 2$-$1 & 	     155.8  &   	166.8  & \phantom{0}8 & &         & &     \\
		 & JCMT  & 3$-$2 & \phantom{0}88.0  & \phantom{0}79.1  & \phantom{0}7 & & \phantom{$<$}10.0\phantom{0} & 9.6\phantom{0} & 7 && \phantom{$>$0}8\phantom{.0} \\
                 & JCMT  & 4$-$3 & \phantom{0}73.3  & \phantom{0}78.6  & \phantom{0}7 & &         & &     \\
\noalign{\smallskip}
\hline
\noalign{\smallskip}
\noalign{1.\ This paper; 2.\ Sch\"oier \& Olofsson (2000, submitted); 3.\ Sch\"oier \& Olofsson (2000, unpublished data); 
4.\ Olofsson et al. \cite*{Olofsson93a};
5.\ Olofsson et al. \cite*{Olofsson93b}; 6.\ Olofsson et al.\ (1996);
7.\ JCMT public archive; 8. Neri et al. \cite*{Neri}; 
9. Nyman et al. \cite*{Nyman93}; 10. Ryde et al. \cite*{Ryde99}.}
\noalign{\smallskip}
\noalign{$^{\mathrm a}$ The observed line ratios have been multiplied by 0.87 to correct for the differences in line 
strenghts and beam-filling factors (see text for details).}
\end{tabular}
\end{flushleft}
\end{table*}

\section{Radiative transfer}
In order to model the circumstellar line emission, and to determine
accurate $^{12}$CO/$^{13}$CO abundance ratios, we have used a non-LTE
radiative transfer code based on the Monte Carlo method [see
Sch\"oier \cite*{Schoeier00} for details].  Assuming a spherically
symmetric CSE expanding at a constant velocity, the code calculates the
molecular excitation, i.e., the level populations, required to solve
the radiative transfer exactly.

The excitation of the CO molecules were calculated taking into account
30 rotational levels in each of the ground and first vibrational
states.  The transition probabilities and energy levels are taken from
Chandra et~al.  \cite*{Chandra96}, and the rotational collisional rate
coefficients (CO-H$_2$) are based on the results in Flower \& Launay
\cite*{Flower85} (they are extrapolated for $J$$>$11 and for
temperatures higher than 250 K).  Collisional transitions between
vibrational levels are not important due to low densities and short
radiative lifetimes.

The basic physical parameters of the CSE, e.g., the mass loss rate,
expansion velocity, and temperature structure, are determined from the
analysis of the observed $^{12}$CO radio line emission.  CO is well
suited for this purpose since it is difficult to photodissociate and
easy to excite through collisions, and thus is a very good tracer of
the molecular gas and its temperature.  The kinetic
temperature of the gas is derived in a self-consistent manner, i.e.,
the calculations include the most important heating and cooling
mechanisms for the gas, e.g., heating due to dust-gas collisions and
molecular line cooling from CO. Once the characteristics of the CSE
have been determined, the $^{13}$CO excitation analysis is performed.
The $^{13}$CO abundance is varied until a satisfactory fit to the
observations is obtained.  In this way the circumstellar
$^{12}$CO/$^{13}$CO-ratio is estimated.

The spatial extent of the molecular envelope is an
important input parameter, and the derived mass loss rate and, to a
lesser extent, the $^{12}$CO/$^{13}$CO-ratio will depend on this.  The size
of the circumstellar CO envelope, assumed to be the same for $^{12}$CO
and $^{13}$CO, was estimated based on the modelling presented in Mamon
et~al.  \cite*{Mamon88}.  It includes photodissociation,
self-shielding and H$_{2}$-shielding, and chemical exchange reactions.
Here we assume the radial fractional abundance distribution to follow
\begin{equation}
f(r) = f_0 \exp \left[ - \ln 2 \left( \frac{r}{r_{\mathrm p}}
\right)^{\alpha} \right],
\end{equation}
where $f_0$ is the initial (photospheric) abundance with respect to H$_{2}$,
$r_{\mathrm p}$ is the photodissociation radius (where the abundance
has dropped to $f_0/2$), and $\alpha$ is a parameter describing the
rate at which the abundance declines.  Both $r_{\mathrm p}$ and
$\alpha$ depend on the mass loss rate, the expansion velocity, and
$f_0$.  When modelling the $^{12}$CO emission we assume
$f_0$$=$1$\times$10$^{-3}$.  This is an average of the $f_0$:s estimated
by Olofsson et~al. \cite*{Olofsson93b} for a sample of optically bright carbon
stars.

In our models we include both a central source of radiation and the
cosmic background radiation at 2.7\,K. The central radiation emanates
from the star itself, which may be approximated by a blackbody.  For
heavily dust-enshrouded objects, like \object{CW~Leo} (a.k.a.\
\object{IRC+10216}), most of the stellar light is re-emitted by dust
at longer wavelenghts.  This emission source is also approximated by a
blackbody.  For low mass loss rate objects, the stellar blackbody
temperature, $T_{*}$, was estimated from a fit to the SED of the object.
For stars of intermediate to high mass loss rates, two blackbodies
were used, one representing the stellar contribution and one
representing the dust.  A fit to the SED gives the two blackbody
temperatures, $T_{*}$ and $T_{\rm d}$, and the relative luminosities
of the two blackbodies, $L_{\rm d}$/$L_{*}$.  The method is described
in Kerschbaum \cite*{Kerschbaum99}.  The temperatures and luminosities
used in the modelling are presented in Table~\ref{input}.  The inner
boundary of the CSE was set to be outside the radius of the central
blackbody(s), but never lower than 1$\times$10$^{14}$\,cm.

The distances, presented in Table~\ref{input}, were estimated using one
of the following methods: the observed Hipparcos parallax, a
period-luminosity relation 
(Groenewegen \& Whitelock 1996\nocite{Groenewegen96}), 
or an assumed bolometric luminosity.
In the two former cases the luminosities were estimated using apparent
bolometric magnitudes and the distances.

In the case of \object{V~Hya} and \object{TX~Psc},
stars with possible bipolar outflows (Heske et~al.\ 1989\nocite{Heske};
Kahane et~al.\ 1996\nocite{Kahane96}), no radiative
transfer analysis was attempted due to the complexity of these
outflows.

\section{Model results}

\subsection{The $^{\it{12}}$CO modelling}
As explained above, the $^{12}$CO line emission is used to derive the basic
parameters of the CSE. The observational constraints for each
source are presented in Table~\ref{coresult}.  The main part of the
observational data used here are the $J$$=$$1$$\rightarrow$$0$
and $J$$=$$2$$\rightarrow$$1$ spectra obtained by Olofsson et~al.
\cite*{Olofsson93a} using the SEST, OSO, and the IRAM 30\,m
telescope at Pico Veleta, Spain (some of these observations have been remade
and the intensities stated in Tab~\ref{coresult} may therefore be somewhat 
different from those originally given in the reference).  
Observations of the two lowest
rotational transitions have also been performed using the NRAO 12\,m
telescope at Kitt Peak, USA, and of the $J$$=$$3$$\rightarrow$$2$
line using the SEST (Sch\"oier \& Olofsson 2000, submitted). In
addition, we have obtained publicly available data from the
James Clerk Maxwell Telescope (JCMT) at Mauna Kea, Hawaii.
The JCMT data are taken at face value. However, in the cases where there are
more than one observation available, the derived line intensities are generally
consistent within $\pm$20\%. In addition, the good agreement with corresponding
SEST observations lend further support for the reliability of the JCMT public
data.

\begin{figure*}
  \resizebox{\hsize}{!}{\includegraphics{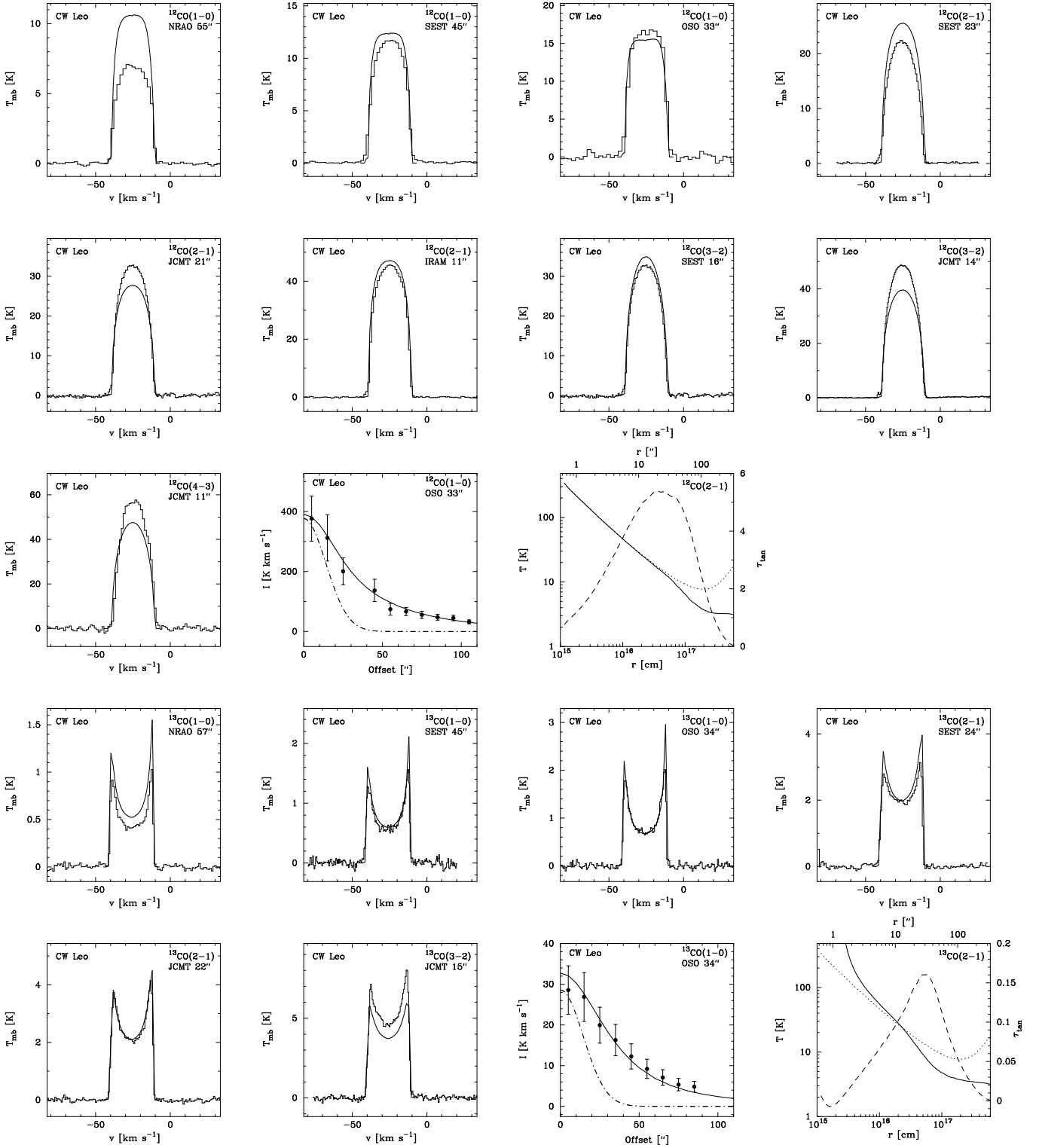}}
  \caption{Multi-transition $^{12}$CO and $^{13}$CO millimetre wave line
  emission observed towards the high mass loss rate Mira variable
  \object{CW~Leo}. The observed spectra (histograms) have been overlayed with
  the model prediction (full line) using a mass loss rate of
  $1.5$$\times$$10$$^{-5}$\,M$_\odot$\,yr$^{-1}$ and a
  $^{12}$CO/$^{13}$CO-ratio of $50$. The transition, telescope used, and
  the beamsize, are given for each of the observations.  Also shown are the
  observed radial brightness distributions overlayed by the results from
  the model (full line).  The circular beam used in the radiative
  transfer is indicated by the dot-dashed line.
  The full lines in the temperature panels give the excitation
  temperature of the $^{12}$CO($J$$=$$2$$\rightarrow$$1$) or the
  $^{13}$CO($J$$=$$2$$\rightarrow$$1$) transition. The dotted lines show
  the kinetic gas temperature derived from the excitation analysis.
  The dashed lines give the tangential optical
  depth, $\tau_{\mbox{tan}}$, of the $^{12}$CO($J$$=$$2$$\rightarrow$$1$) or
  the $^{13}$CO($J$$=$$2$$\rightarrow$$1$) transition}
  \label{cwleo}
\end{figure*}

\begin{figure*}
  \resizebox{\hsize}{!}{\includegraphics{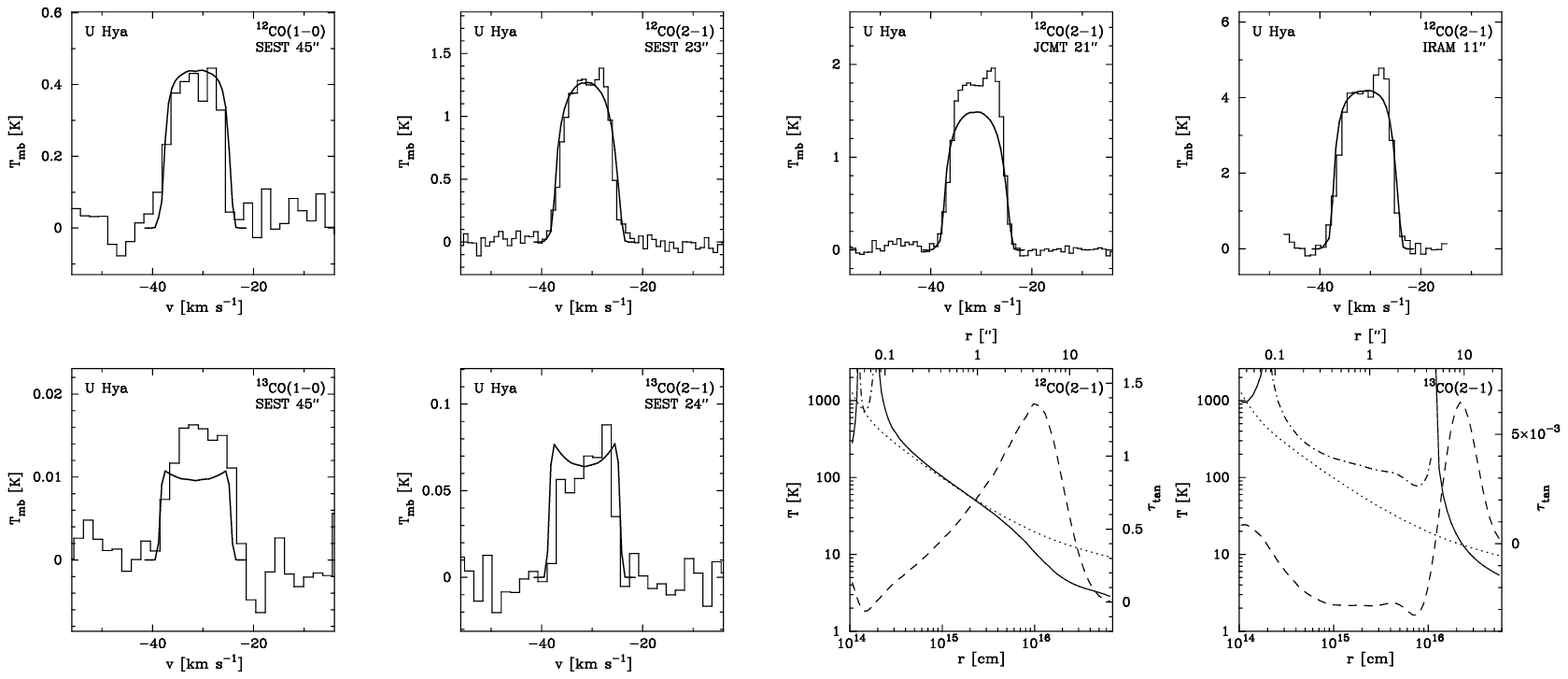}}
  \caption{Multi-transition $^{12}$CO and $^{13}$CO millimetre wave line
  emission observed towards the optically bright carbon star
  \object{U~Hya}. The observed spectra (histograms) have been overlayed with
  the model prediction (full line) using a mass loss rate of
  $1.0$$\times$$10$$^{-7}$\,M$_\odot$\,yr$^{-1}$ and a
  $^{12}$CO/$^{13}$CO-ratio of $25$.
  The transition, telescope used, and
  the beamsize, are given for each of the observations.
  The full lines in the temperature panels give the excitation
  temperature of the $^{12}$CO($J$$=$$2$$\rightarrow$$1$) or the
  $^{13}$CO($J$$=$$2$$\rightarrow$$1$) transition, dash-dot lines indicate a
  negative excitation temperature, i.e., maser action. The dotted lines show
  the kinetic gas temperature derived from the excitation analysis.
  The dashed lines give the tangential optical
  depth, $\tau_{\mbox{tan}}$, of the $^{12}$CO($J$$=$$2$$\rightarrow$$1$) or
  the $^{13}$CO($J$$=$$2$$\rightarrow$$1$) transition}
  \label{uhya}
\end{figure*}

Thus, for all stars we use data from more than one $^{12}$CO
transition, in some cases four, to constrain the model.  The
intensities and overall line shapes of the circumstellar lines
produced by the radiative transfer model generally agree well with
those observed.  This is illustrated here for two of our sample stars
\object{CW~Leo} (Fig.~\ref{cwleo}) and \object{U~Hya}
(Fig.~\ref{uhya}).  \object{CW~Leo} is a high mass loss rate Mira
variable where the excitation of $^{12}$CO is dominated by collisions.
\object{U~Hya}, on the other hand, is a low mass loss rate object
where radiation emitted by the central star plays a role in the
excitation.  The $^{12}$CO model results presented in this paper
constitute a sub-sample of the results of an analysis of a large
survey of carbon stars presented in Sch\"oier \& Olofsson (2000, submitted).  
The reader is referred to this paper for a detailed
description of the sensitivity of the model to the various input
parameters.  However, we point out here that tests made by Sch\"oier
\& Olofsson show that the derived mass loss rate, for the majority of objects 
in this study, is mostly affected by the temperature structure and not
by the assumed inner radius of the shell or the luminosity of the star, i.e.,
collisional excitation dominates over radiative excitation.
In addition, Sch\"oier \& Olofsson
have tested the molecular envelope size calculations by comparing radial
brightness distributions obtained from the model with those observed.
It was found that the envelope sizes estimated using the results from
the model by Mamon et~al.  \cite*{Mamon88} are generally consistent
with the observations.  This is illustrated here in Fig.~\ref{cwleo}
for \object{CW~Leo} where the radial brightness distribution of the
$^{12}$CO($J$$=$$1$$\rightarrow$$0$) line emission, observed at OSO,
is compared to the distribution obtained from the radiative transfer
model.

The derived envelope parameters, used as input for the $^{13}$CO
modelling, are presented in Table~\ref{input}.  We believe that, within
the adopted circumstellar model, the estimated mass loss rates are
generally accurate to within $\pm$50\% (neglecting errors introduced by the
uncertain CO abundance and the distance estimates). In the models presented
here only $^{12}$CO line cooling was included. For the J-stars, cooling from
$^{13}$CO will be important (scales with the isotope ratio for these low
mass loss rate objects), but it will not affect the derived mass loss
rate since the heating must be increased, i.e., the kinetic temperature
structure will not change significantly, in order to maintain a good
fit (Sch\"oier \& Olofsson 2000, submitted).  See also the discussion
in Ryde et~al.  \cite*{Ryde99} for the high mass loss rate object
\object{IRAS~15194-5115}, where radio and far-IR $^{12}$CO and
$^{13}$CO data are modelled.

When comparing our mass loss rate estimate for \object{CW~Leo} to those
obtained from other detailed radiative transfer models (Kastner
1992\nocite{Kastner92}; Crosas \& Menten 1997\nocite{Crosas97};
Groenewegen et~al.\ 1998\nocite{Groenewegen98}) we find a very good
agreement, within 20\%, when adjustments for differences in $f_0$ and
distance have been made.  Kastner \cite*{Kastner92} also modelled the
high mass loss rate object \object{RW~LMi} (a.k.a. \object{CIT~6})
obtaining (when corrected for the difference in distance) a mass loss
rate in excellent agreement with our estimate.  Sopka et~al.
\cite*{Sopka89} used a much simpler radiative transfer model to derive
the mass loss rate for a number of AGB stars, of which four overlap
with our survey.  For \object{RW~LMi} they derived a mass loss rate in
excellent agreement with that obtained from our model.  For 
\object{LP~And}, \object{V384~Per}, and \object{V~Cyg}, however, there are
discrepancies of about a factor of two to three.  It should be noted
though that Sopka et~al.  \cite*{Sopka89} based their mass loss rate
estimates on observations of a single line, $J$$=$$1$$\rightarrow$$0$,
using a single telescope (OSO). For \object{LP~And}, the source with the
largest
discrepancy, we observe a $^{12}$CO($J$$=$$1$$\rightarrow$$0$) line
that is almost a factor of two stronger using the same telescope.

\subsection{The $^{\it{13}}$CO modelling}

Once the general characteristics of the CSE have been determined from
the $^{12}$CO analysis, the observed $^{13}$CO line emission is
modelled.  Again, the observed intensities, as well as the line
shapes, are generally well reproduced by the model.  As for the
$^{12}$CO modelling, this is illustrated for two of our sample stars,
\object{CW~Leo} (Fig.~\ref{cwleo}) and \object{U~Hya}
(Fig.~\ref{uhya}).

The abundance of $^{13}$CO is usually (the J-stars provide an
exception) much smaller than that of $^{12}$CO, which leads to
significantly different excitation conditions for the rarer
isotopomer.  For instance, the model $^{13}$CO line intensities are
more sensitive to the assumed properties of the central source of
emission, since radiative pumping via the first excited vibrational
state becomes important.  Indeed, the $^{13}$CO
$J$$=$$1$$\rightarrow$$0$ and $J$$=$$2$$\rightarrow$$1$ transitions
have inverted populations over parts of the envelope even for a high
mass loss rate object as \object{CW~Leo}, Fig.~\ref{cwleo}.  For the
high mass loss rate objects, however, the part over which there is a
weak maser acting is very small and the emission emanating from this
region is not detected in our observations.  For thinner envelopes the
part of the envelope where the lowest rotational levels are inverted
is larger due to the fact that the pumping emission can penetrate
further out into the wind.  In the case of \object{U~Hya}, changing
the inner radius of the CSE or the luminosity of the star by
$\sim$50\% will change the estimated $^{13}$CO abundance by
$\sim$20\%.

In our modelling of the $^{13}$CO line emission we have assumed that the
$^{13}$CO envelope size is equal to that of $^{12}$CO. This is based
upon the model results by Mamon et~al. \cite*{Mamon88}.  In this
model both the effects of photodissociation and of chemical
fractionation were included.  
 Chemical fractionation of CO occurs through the exchange reaction
(Watson et al.\ 1976\nocite{Watson76})
$^{13}{\mathrm C}^+$$+$$^{12}{\mathrm{CO}}\,\leftrightarrows\,^{12}{\mathrm C}^+$$+$$^{13}{\mathrm{CO}}$$+$$\Delta {\mathrm E}$,  
where $\Delta {\mathrm E}/k$$=$35\,K. Below this temperature, the backward 
reaction is suppressed and production of $^{13}$CO from $^{12}$CO is favoured.
This reaction  can effectively produce
$^{13}$CO in the outer, cool parts of circumstellar envelopes.
Mamon et~al. \cite*{Mamon88} concluded that the difference
between the $^{12}$CO and $^{13}$CO abundance
distributions, when tested over a large mass loss rate interval, is
always small, no more than 10 to 20\%.  Without the effect of
chemical fractionation the $^{13}$CO envelope would be
significantly smaller.  This is due to the fact that CO is
photodissociated in lines and thereby exhibits considerable
self-shielding.  Thus, the shielding is higher the higher the optical
depth, and hence it is less efficient for the less abundant $^{13}$CO.
Using OSO we have obtained a brightness distribution map of the
$^{13}$CO($J$$=$$1$$\rightarrow$$0$) emission around \object{CW~Leo}.
It is found that the model, with an assumed $^{13}$CO envelope size
equal to that of $^{12}$CO, reproduces the observed radial brightness
distribution within the observational errors, Fig.~\ref{cwleo}.
A 20\% smaller $^{13}$CO envelope gives the same
$^{12}$CO/$^{13}$CO-ratio but fails to reproduce the observed radial
brightness distribution within the observational errors.  This shows
the importance of chemical fractionation, at least in the cool outer
parts of dense CSEs.  In the case of the thin molecular envelope
around \object{U~Hya} a 20\% smaller $^{13}$CO envelope size would
increase the derived $^{13}$CO abundance by about 20\%. 
 This
illustrates that in the case of a thin CSE the CO molecules are effectively
excited to the photodissociation radius, i.e., photodissociation determines
the size of the emitting region, whereas in a dense CSE excitation sets
the size of the emitting region.

\begin{figure}
  \resizebox{\hsize}{!}{\includegraphics{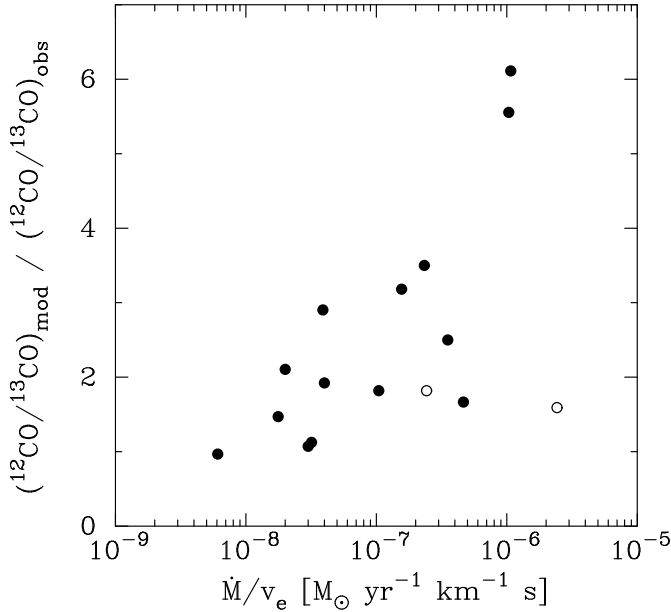}}
  \caption{The ratio between the $^{12}$CO/$^{13}$CO-ratio obtained
  from the radiative transfer analysis and that obtained from the observed
  line intensity ratios (in the cases where more than one value is available in Table~\ref{coresult} 
  we have used an average)
  as a function of the density of the shell ($\sim$$\dot{M}$/$v_{\mathrm e}$).
  Sources with detached CSEs are marked as open circles.}
  \label{ratio}
\end{figure}

\subsection{The circumstellar $^{\it{12}}$CO/$^{\it{13}}$CO-ratio}

In Table~\ref{coresult} we list the observed integrated
$^{12}$CO/$^{13}$CO line intensity ratios. They are corrected for the
differences in line strengths and beam-filling factors (assuming the
sources to be unresolved the combined effect gives a $\nu^{-3}$ correction,
i.e., the observed ratio is lowered by a factor of 0.87). 
Although we note that some of our CSEs are resolved, we have nevertheless 
treated all our sources in the same manner.  
Here we have chosen to integrate the emission over
the entire line in order to achieve better signal-to-noise ratios.
For optically thin lines this ratio should provide a first order
estimate of the abundance ratio.  A narrow velocity interval centered
on the systemic velocity, where optical depth effects are smallest,
should give somewhat higher line intensity ratios than those presented
here. Nine of our sample stars have been detected in the
$^{13}$CO($J$$=$$1$$\rightarrow$$0$) and/or the
$^{13}$CO($J$$=$$2$$\rightarrow$$1$) line.  A simple comparison of the
$^{12}$CO/$^{13}$CO line intensity ratios and the
$^{12}$C/$^{13}$C-ratios derived by Lambert et~al.  \cite*{Lambert86}
shows a tight correlation of the form (in the cases where more than one value
is available in Table~\ref{coresult} we have used an average),
\begin{equation}
\frac{I(^{12}\mathrm{CO})}{I(^{13}\mathrm{CO})} = (0.6\pm 0.2)
\times \frac{^{12}\mathrm{C}}{^{13}\mathrm{C}}.
\end{equation}
Thus, a straightforward use of line intensity ratios would lead to
$^{12}$C/$^{13}$C-ratios that, on the average, agree with those of Ohnaka
\& Tsuji \cite*{Ohnaka96}.  However, any optical depth effects would
lower the observed line intensity ratio.

In Table~\ref{input} we present the  $^{12}$CO/$^{13}$CO-ratios
derived using our radiative transfer code.  They span a large range,
from 2.5 to 90 
 (steps of 5 in the $^{12}$CO/$^{13}$CO-ratio was used to
find the best fit $^{13}$CO model, except for the
J-stars where a smaller step-size of 0.5 was used).
For most of our observed stars the isotope ratio obtained
from the detailed radiative transfer is higher than those estimated
from simple line intensity ratios.  The discrepancy is fairly small
for the objects with thin CSEs, but it increases with the thickness of
the CSE ($\sim$$\dot{M}$/$v_{\mathrm e}$) and reaches a factor of six for the
high mass loss rate objects, Fig.~\ref{ratio}.
This reflects that even for low to intermediate mass loss
rate objects there are optical depth effects, and a detailed modelling
is needed to derive reliable isotope ratios.  In Table~\ref{input} we
present, for each star, the maximum tangential optical depth in the
$^{12}$CO($J$$=$$2$$\rightarrow$$1$) transition obtained in the
modelling.  The radial variation of the tangential optical depth is
shown for the high mass loss rate object \object{CW~Leo} in
Fig.~\ref{cwleo} and for the low mass loss rate object \object{U~Hya}
in Fig.~\ref{uhya}.  The optical depth in the
$^{12}$CO($J$$=$$1$$\rightarrow$$0$) line is significantly lower.

The estimated $^{13}$CO abundance depends on the adopted $^{12}$CO
abundance (assumed to be equal for all stars).  We find that to a
first approximation it scales with $f_{0}$, since a lower (higher)
$f_{0}$ leads to a higher (lower) mass loss rate to fit the
$^{12}$CO line intensities and hence a lower (higher) $^{13}$CO abundance
to fit the $^{13}$CO line intensities.  This means that to a first
approximation the estimated $^{12}$CO/$^{13}$CO-ratios are only very
weakly dependant on the adopted $^{12}$CO abundance. We have varied
some of the other parameters in the model (see above), and conclude that
in doing so the derived $^{12}$CO/$^{13}$CO-ratio changes by about 20\%.  
With an additional uncertainty of $\pm$15\% in the
relative calibration of the $^{12}$CO and $^{13}$CO data
 (the relative calibration is usually better than the absolute 
calibration for any given telescope)
, we estimate
that the $^{12}$CO/$^{13}$CO-ratio is uncertain by about $\pm$30\%.
In the cases were the $^{12}$CO/$^{13}$CO-ratio estimate relies on observations
of just one line the uncertainty may be as high as $\pm$50\% depending on 
the quality of the data.

We compare first our derived $^{12}$CO/$^{13}$CO-ratio for \object{CW~Leo}
with those found by others. Crosas \& Menten \cite*{Crosas97} derived,
using a radiative transfer model similar to ours, an isotope ratio of 50, i.e.,
the same as we do.  Greaves \& Holland \cite*{Greaves97} found a lower
ratio of 32, and also a much lower ratio (24) for \object{LP~And}, using
a simple radiative transfer model.  Kahane et~al.  \cite*{Kahane92}
found the $^{12}$C/$^{13}$C-ratio to be 44 based on observations of
optically thin lines.  Kahane et~al.  \cite*{Kahane92} also determined
the $^{12}$C/$^{13}$C-ratio for \object{RW~LMi} obtaining a value of
31, in good agreement with our $^{12}$CO/$^{13}$CO estimate for this
object.  Sopka et~al.  \cite*{Sopka89} estimated
$^{12}$CO/$^{13}$CO-ratios for four of our stars that are all lower
than our derived values.  This is most probably an effect of the
difference in the treatment of the radiative transfer.
We also note that Dufour et~al.\ (2000, in prep.), 
derive somewhat lower $^{12}$CO/$^{13}$CO-ratios for the three J-stars.
 
\begin{figure}
  \resizebox{\hsize}{!}{\includegraphics{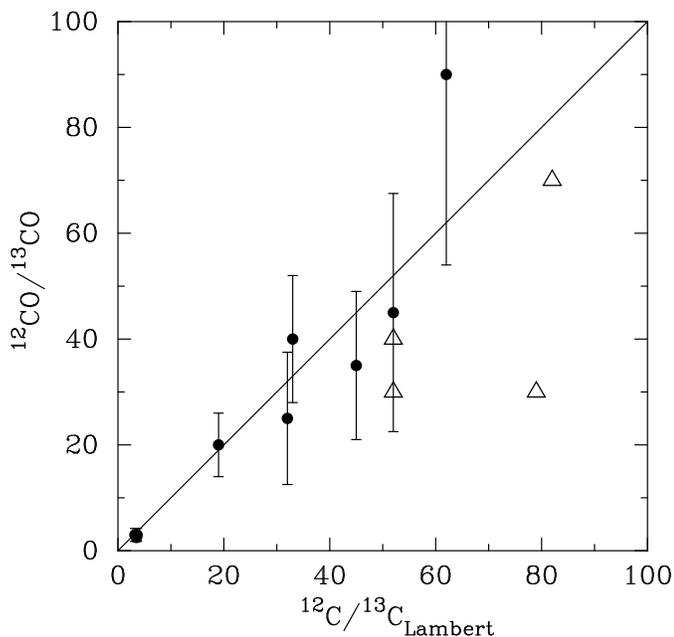}}
  \caption{Comparison between the estimated circumstellar
  $^{12}$CO/$^{13}$CO-ratio obtained using our detailed radiative
  transfer code, and the photospheric $^{12}$C/$^{13}$C-ratio estimated
  by Lambert et~al.\ (1986).  An open triangle indicate a lower limit and the
  solid line represents a 1:1 correlation.  Note that at
  $^{12}$C/$^{13}$C$\sim$3 there are results for three stars.}
  \label{ResPlot}
\end{figure}

For our sample stars we compare the circumstellar
$^{12}$CO/$^{13}$CO-ratios obtained from the modelling with the
photospheric $^{12}$C/$^{13}$C-ratios estimated by Lambert et~al.
\cite*{Lambert86} in Fig~\ref{ResPlot}.  A good correlation of the form
\begin{equation}
\frac{^{12}\mathrm{CO}}{^{13}\mathrm{CO}} = (1.0\pm 0.2) \times
\frac{^{12}\mathrm{C}}{^{13}\mathrm{C}}
\end{equation}
is obtained. Lambert et~al. \cite*{Lambert86} give an uncertainty in their
estimated $^{12}$C/$^{13}$C-ratios of about $\pm$40\%.
Included in our sample are stars that have had a drastic change 
in their mass loss rate.  For \object{R~Scl} and \object{S~Sct}, stars
with known detached CSEs (probably) produced during a period of
intense mass loss, we have performed the analysis using the envelope
parameters determined by Olofsson et~al.  \cite*{OBEG}.  These results
are included in Fig~\ref{ResPlot}.  We note here that our results
suggest that the $^{12}$C/$^{13}$C-ratios in the detached shells (with
ages of about 10$^3$ and 10$^4$\,yr for \object{R~Scl} and
\object{S~Sct}, respectively) are the same as the present ones in the
photospheres.  

\section{The $^{12}$C/$^{13}$C-ratio}

Taken at face values our derived $^{12}$CO/$^{13}$CO-ratios support
the $^{12}$C/$^{13}$C-ratios derived by Lambert et~al. \cite*{Lambert86},
Fig~\ref{ResPlot}.  However, there are a number of uncertainties in
the extrapolation from the circumstellar isotopomer result to the
stellar isotope ratio.  The properties of the CO molecule suggests
that the isotopomer ratio in the gas leaving the star is equal to the
isotope ratio.  We have based our circumstellar model upon the
photodissociation/chemical fractionation results of Mamon et~al.
\cite*{Mamon88}, and this has so far proven to give good results in
our test cases.  Our radiative transfer calculations are detailed and
also provide good fits to multi-transition data.  We also believe that
estimates of isotopomer ratios are far less dependant on the adopted
circumstellar model than are individual abundances.  
 The two molecules have essentially the same energy level diagrams,
the same transition strengths, and the same collisional cross sections, and
hence their relative abundances are much less dependant on the adopted model
than their absolute abundances (note that for $^{12}$CO we actually 
adopt an abundance).
However, one
should note that the circumstellar estimates apply to time scales of
10$^2$ to 10$^3$ years, but there is no reason to expect that these
stars have changed their surface composition over such a short time
scale.  In conclusion, we believe that our derived
$^{12}$CO/$^{13}$CO-ratios are reliable estimates of the stellar
$^{12}$C/$^{13}$C-ratios.  Thus, we support the results obtained by
Lambert et~al.  \cite*{Lambert86}.

Also for the J-stars, whose origin is uncertain, we estimate
$^{12}$C/$^{13}$C-ratios that are more consistent with those of
Lambert et~al.  \cite*{Lambert86} than those of Ohnaka \& Tsuji
(1999)\nocite{Ohnaka99}, i.e., ratios that are, at least in principle,
possible to obtain within the CNO-cycle.  \object{IRAS~15194-5115}
also has a low $^{12}$C/$^{13}$C-ratio, but it differs from the
J-stars in the sense that it has a much higher mass loss rate (see
also Ryde et~al. 1999\nocite{Ryde99}).  This star could be a borderline
case with a mass of about 3.5\,M$_{\sun}$, where the CNO-cycle produces
a low $^{12}$C/$^{13}$C-ratio, while the temperature is not high enough to
effectively convert  $^{12}$C to $^{14}$N 
(Ventura et~al.\ 1999\nocite{Ventura99}).

\section{Conclusions}

We have determined the $^{12}$CO/$^{13}$CO-ratio in the molecular envelopes
around 20 carbon stars, using a detailed non-LTE radiative transfer code.
The $^{12}$CO/$^{13}$CO-ratios found range from 2.5 to 90, and we
believe that these ratios accurately measure the stellar
$^{12}$C/$^{13}$C-ratios.

Due to optical depth effects, present mainly in the $^{12}$CO line, we
find it necessary to treat in detail the radiative transfer in order to
obtain reliable isotopomer ratios.  For instance, for the two stars in
common, we derive $^{12}$CO/$^{13}$CO-ratios that are almost a factor
of two higher than those of Greaves \& Holland \cite*{Greaves97} in
their study of the evolution of the local interstellar
$^{12}$C/$^{13}$C-ratio.

An important ingredient in the model is the molecular abundance
distribution in the CSE. We have here used the results of Mamon et~al.
\cite*{Mamon88} for $^{12}$CO and $^{13}$CO, and we find that for both
molecules these are consistent with the observations, although for
$^{13}$CO we have only been able to test the results for a high mass
loss rate object.

Our estimated $^{12}$C/$^{13}$C-ratios agree well with those estimated
in the photosphere by Lambert et~al. \cite*{Lambert86}.  Thus, we do
not support the claims by other authors that the isotope ratios derived by
Lambert et~al. \cite*{Lambert86} were too high, by about a factor of two.

\begin{acknowledgements}
      We are grateful to Dr.~F.~Kerschbaum for generously providing
      estimates of some of the input parameters to the CO modelling.
      We would also like to thank Dr.~C.~Kahane for valuable comments.
      Financial support from the Swedish Natural Science Research Council
      (NFR) is gratefully acknowledged.
\end{acknowledgements}


\begin{thebibliography}{}

\bibitem[\protect\astroncite{{Abia} and {Isern}}{1997}]{Abia97}
{Abia} C., {Isern} J., 1997,
\newblock {MNRAS} {289}, L11

\bibitem[\protect\astroncite{{Bergman} et~al.}{1993}]{Bergman93}
{Bergman} P., {Carlstr\"{o}m} U., {Olofsson} H., 1993,
\newblock {A\&A} {268}, 685

\bibitem[\protect\astroncite{{Boothroyd} et~al.}{1993}]{Boothroyd93}
{Boothroyd} A.~I., {Sackmann} I.~J., {Ahern} S.~C., 1993,
\newblock {ApJ} {416}, 762

\bibitem[\protect\astroncite{{Busso} et~al.}{1999}]{Busso99}
{Busso} M., {Gallino} R., {Wasserburg} G.~J., 1999,
\newblock {ARA\&A} {37}, 239

\bibitem[\protect\astroncite{{Chandra} et~al.}{1996}]{Chandra96}
{Chandra} S., {Maheshwari} V., {Sharma} A., 1996,
\newblock {A\&AS} {117}, 557

\bibitem[\protect\astroncite{{Crosas} and {Menten}}{1997}]{Crosas97}
{Crosas} M., {Menten} K.~M., 1997,
\newblock {ApJ} {483}, 913

\bibitem[\protect\astroncite{{de Laverny} and {Gustafsson}}{1998}]{deLaverny98}
{de Laverny} P., {Gustafsson} B., 1998,
\newblock {A\&A} {332}, 661

\bibitem[\protect\astroncite{{de Laverny} and {Gustafsson}}{1999}]{deLaverny99}
{de Laverny} P., {Gustafsson} B., 1999,
\newblock {A\&A} {346}, 520

\bibitem[\protect\astroncite{{Dominy} et~al.}{1986}]{Dominy86}
{Dominy} J.~F., {Wallerstein} G., {Suntzeff} N.~B., 1986,
\newblock {ApJ} {300}, 325

\bibitem[\protect\astroncite{{Flower} and {Launay}}{1985}]{Flower85}
{Flower} D.~R., {Launay} J.~M., 1985,
\newblock {MNRAS} {214}, 271

\bibitem[\protect\astroncite{{Forestini} and {Charbonnel}}{1997}]{Forestini97}
{Forestini} M., {Charbonnel} C., 1997,
\newblock {A\&AS} {123}, 241

\bibitem[\protect\astroncite{{Greaves} and {Holland}}{1997}]{Greaves97}
{Greaves} J.~S., {Holland} W.~S., 1997,
\newblock {A\&A} {327}, 342

\bibitem[\protect\astroncite{{Groenewegen} and
  {Whitelock}}{1996}]{Groenewegen96}
{Groenewegen} M. A.~T., {Whitelock} P.~A., 1996,
\newblock {MNRAS} {281}, 1347

\bibitem[\protect\astroncite{{Groenewegen} et~al.}{1998}]{Groenewegen98}
{Groenewegen} M. A.~T., {van Der Veen} W. E. C.~J., {Matthews} H.~E., 1998,
\newblock {A\&A} {338}, 491

\bibitem[\protect\astroncite{{Heske} et~al.}{1989}]{Heske}
{Heske} A., {Te Lintel Hekkert} P., {Maloney} P.~R., 1989,
\newblock {A\&A} {218}, L5

\bibitem[\protect\astroncite{{Kahane} et~al.}{1992}]{Kahane92}
{Kahane} C., {Cernicharo} J., {Gomez-Gonzalez} J., {Guelin} M., 1992,
\newblock {A\&A} {256}, 235

\bibitem[\protect\astroncite{{Kahane} et~al.}{1996}]{Kahane96}
{Kahane} C., {Audinos} P., {Barnbaum} C., {Morris} M., 1996,
\newblock {A\&A} {314}, 871

\bibitem[\protect\astroncite{{Kastner}}{1992}]{Kastner92}
{Kastner} J.~H., 1992,
\newblock {ApJ} {401}, 337

\bibitem[\protect\astroncite{{Kerschbaum}}{1999}]{Kerschbaum99}
{Kerschbaum} F., 1999,
\newblock {A\&A} {351}, 627

\bibitem[\protect\astroncite{{Knapp} and {Chang}}{1985}]{Knapp85}
{Knapp} G.~R., {Chang} K.~M., 1985,
\newblock {ApJ} {293}, 281

\bibitem[\protect\astroncite{{Lambert} et~al.}{1986}]{Lambert86}
{Lambert} D.~L., {Gustafsson} B., {Eriksson} K., {Hinkle} K.~H., 1986,
\newblock {ApJS} {62}, 373

\bibitem[\protect\astroncite{{Mamon} et~al.}{1988}]{Mamon88}
{Mamon} G.~A., {Glassgold} A.~E., {Huggins} P.~J., 1988,
\newblock {ApJ} {328}, 797

\bibitem[\protect\astroncite{{Neri} et~al.}{1998}]{Neri}
{Neri} R., {Kahane} C., {Lucas} R., {Bujarrabal} V., {Loup} C., 1998,
\newblock {A\&AS} {130}, 1

\bibitem[\protect\astroncite{{Nyman} et~al.}{1993}]{Nyman93}
{Nyman} L.~A., {Olofsson} H., {Johansson} L. E.~B., et~al., 1993,
\newblock {A\&A} {269}, 377

\bibitem[\protect\astroncite{{Ohnaka} and {Tsuji}}{1996}]{Ohnaka96}
{Ohnaka} K., {Tsuji} T., 1996,
\newblock {A\&A} {310}, 933

\bibitem[\protect\astroncite{{Ohnaka} and {Tsuji}}{1998}]{Ohnaka98}
{Ohnaka} K., {Tsuji} T., 1998,
\newblock {A\&A} {335}, 1018

\bibitem[\protect\astroncite{{Ohnaka} and {Tsuji}}{1999}]{Ohnaka99}
{Ohnaka} K., {Tsuji} T., 1999,
\newblock {A\&A} {345}, 233

\bibitem[\protect\astroncite{{Ohnaka} et~al.}{2000}]{Ohnaka00}
{Ohnaka} K., {Tsuji} T., {Aoki} W., 2000,
\newblock {A\&A} {353}, 528

\bibitem[\protect\astroncite{{Olofsson} et~al.}{1993a}]{Olofsson93a}
{Olofsson} H., {Eriksson} K., {Gustafsson} B., {Carlstr\"om} U., 1993a,
\newblock {ApJS} {87}, 267

\bibitem[\protect\astroncite{{Olofsson} et~al.}{1993b}]{Olofsson93b}
{Olofsson} H., {Eriksson} K., {Gustafsson} B., {Carlstr\"om} U., 1993b,
\newblock {ApJS} {87}, 305

\bibitem[\protect\astroncite{{Olofsson} et~al.}{1996}]{OBEG}
{Olofsson} H., {Bergman} P., {Eriksson} K., {Gustafsson} B., 1996,
\newblock {A\&A} {311}, 587

\bibitem[\protect\astroncite{{Prantzos} et~al.}{1996}]{Prantzos96}
{Prantzos} N., {Aubert} O., {Audouze} J., 1996,
\newblock {A\&A} {309}, 760

\bibitem[\protect\astroncite{Ryde et~al.}{1999}]{Ryde99}
Ryde N., Sch\"{o}ier F.~L., Olofsson H., 1999,
\newblock {A\&A} {345}, 841

\bibitem[\protect\astroncite{Sch\"oier}{2000}]{Schoeier00}
Sch\"{o}ier F.~L., 2000, PhD thesis, Stockholm Observatory

\bibitem[\protect\astroncite{{Smith} and {Lambert}}{1985}]{Smith85}
{Smith} V.~V., {Lambert} D.~L., 1985,
\newblock {ApJ} {294}, 326

\bibitem[\protect\astroncite{{Smith} and {Lambert}}{1990}]{Smith90}
{Smith} V.~V., {Lambert} D.~L., 1990,
\newblock {ApJS} {72}, 387

\bibitem[\protect\astroncite{{Sopka} et~al.}{1989}]{Sopka89}
{Sopka} R.~J., {Olofsson} H., {Johansson} L. E.~B., {Nguyen} Q.-R., {Zuckerman}
  B., 1989,
\newblock {A\&A} {210}, 78

\bibitem[\protect\astroncite{{Ventura} et~al.}{1999}]{Ventura99}
{Ventura} P., {D'Antona} F., {Mazzitelli} I., 1999,
\newblock {ApJ Lett.} {524}, L111

\bibitem[\protect\astroncite{{Wallerstein} and {Knapp}}{1998}]{WK}
{Wallerstein} G., {Knapp} G.~R., 1998,
\newblock {ARA\&A} {36}, 369

\bibitem[\protect\astroncite{{Watson} et~al.}{1976}]{Watson76}
{Watson} W.~D., {Anicich} V.~G., {Huntress} W.~T. J., 1976,
\newblock {ApJ Lett.} {205}, L165

\end{thebibliography}
\bibliographystyle{aabib99}

\end{document}